# An instrumented hammer to detect the bone transitions during an high tibial osteotomy: an animal study

**Bas dit Nugues Manon, first author**
CNRS, Univ Paris Est Creteil, Univ Gustave Eiffel, UMR 8208, MSME,
61 Avenue du Général de Gaulle, 94010 Créteil Cedex, France
manon.bas-dit-nugues@u-pec.fr

**Teddy Ketani, second author**
CNRS, Univ Paris Est Creteil, Univ Gustave Eiffel, UMR 8208, MSME,
61 Avenue du Général de Gaulle, 94010 Créteil Cedex, France
teddy.ketani-mayindou@u-pec.fr

**Claire Bastard, third author**
Sorbonne Université, Paris, France. Service de Chirurgie de la Main, Service
d'Orthopédie et de Traumatologique, Hôpital Saint-Antoine, Sorbonne Université
184, rue du Faubourg-Saint-Antoine, 75012 Paris, France
claire.bastard@aphp.fr

**Giuseppe Rosi, fourth author**
Univ Paris Est Creteil, Univ Gustave Eiffel, CNRS, UMR 8208, MSME,
61 Avenue du Général de Gaulle, 94010 Créteil Cedex, France
giuseppe.rosi@u-pec.fr

**Hugues Albini Lomani, fifth author**
CNRS, Univ Paris Est Creteil, Univ Gustave Eiffel, UMR 8208, MSME,
61 Avenue du Général de Gaulle, 94010 Créteil Cedex, France
hugues.albini@gmail.com

**Charles Henri Flouzat-Lachaniette, sixth author**
INSERM U955, IMRB Université Paris-Est / Service de Chirurgie Orthopédique et
Traumatologique, Hôpital Henri Mondor AP-HP, CHU Paris 12, Université Paris-Est,
51 avenue du Maréchal de Lattre de Tassigny, 94000 Créteil, France
charles-henri.flouzat-lachaniette@aphp.fr

**Arnaud Dubory, seventh author**
INSERM U955, IMRB Université Paris-Est / Service de Chirurgie Orthopédique et
Traumatologique, Hôpital Henri Mondor AP-HP, CHU Paris 12, Université Paris-Est,
51 avenue du Maréchal de Lattre de Tassigny, 94000 Créteil, France
arnaud.dubory@aphp.fr

**Guillaume Haïat, eighth author**[1]

Affiliation

61 Avenue du Général de Gaulle, 94010 Créteil Cedex, France

guillaume.haiat@cnrs.fr

[1] Corresponding author

## ABSTRACT

*High Tibial Osteotomy is a common procedure for knee osteoarthritis during which the surgeon partially opens the tibia and must stop impacting when cortical bone is reached by the osteotome. Surgeons rely on their proprioception and fluoroscopy to conduct the surgery. Our group has developed an instrumented hammer to assess the mechanical properties of the material surrounding the osteotome tip. The aim of this* ex vivo *study is to determine whether this hammer can be used to detect the transition from cortical to trabecular bone and vice versa.*

*Osteotomies were performed until rupture in pig tibia using the instrumented hammer. An algorithm was developed to detect both transitions based on the relative variation of an indicator derived from the time variation of the force. The detection by the algorithm of both transitions was compared with the position of the osteotome measured with a video camera and with surgeon proprioception.*

*The difference between the detection of the video and the algorithm (respectively the video and the surgeon; the surgeon and the algorithm) is 1.0 ± 1.5 impacts (respectively 0.5 ± 0.6 impacts; 1.4 ± 1.8 impacts), for the detection of the transition from the cortical to trabecular bone. For the transition from the trabecular to cortical bone, the difference is 3.6 ± 2.6 impacts (respectively 3.9 ± 2.4 impacts; 0.8 ± 0.9 impacts) and the detection by the algorithm was always done before the sample rupture.*

*This* ex vivo *study demonstrates that this method could prevent impacts leading to hinge rupture.*

*250 Words*

## 1. INTRODUCTION

High Tibial Osteotomies (HTO) are an effective alternative to knee arthroplasty [1,2] to treat osteoarthritis [2,3] in particular for young and active patients [1,4]. Osteotomies are common procedures where a surgeon uses a tool with a sharp blade (an osteotome), which is impacted with a surgical mallet to cut the bone. During an HTO, the surgeon needs to cut the tibia in order to re-align the tibial plateau and re-distribute stresses in a balanced manner [5], which allows to i) reduce the pain and ii) postpone possible knee arthroplasty [5,6]. During an open-wedge HTO the surgeon creates a hinge in the tibia with a sufficient angle of correction to straighten the tibial plateau. The space created in the bone is then filled with an appropriate bone filler and both bone parts are then tightened with plates and screws (see Fig. 1). During this procedure, the tip of the osteotome is moved in a plane perpendicular to the bone axis first across cortical bone, and it then reaches trabecular bone. The osteotomy procedure should continue until the osteotome reaches the second (opposite) cortical bone layer. The surgeon should then stop impacting in order to keep a solid hinge, that is constituted by the cortical layer. Therefore, the detection of the moment when the tip of the osteotome crosses different kinds of bone tissue is important to avoid the occurrence of broken hinge, which requires additional surgical procedures and induces post-operative pain and longer recovery time for the patient [7,8]. In case of a broken cortical hinge, another consequence could also be the damage of the surrounding soft tissues if the osteotome tip goes out of the second cortical bone layer.

Like most osteotomy procedures, HTO is performed with little or no visual control [6]. Here, the open-wedge HTO procedure is carried out through an incision of 5 mm in order to minimize scars on the patient. This surgical procedure is based on the surgeon proprioception and in particular on the sound produced by the impacts of the hammer on the osteotome.

Different solutions have been proposed in the literature to assist the surgeon during osteotomies. The surgeon can use fluoroscopy, an X-ray imaging method, to help position guides allowing to locate the osteotome during the procedure [6]. Other tools such as CT-based intraoperative navigation systems [9] or CT-free navigation systems [10,11], based on fluoroscopic images, have been developed for HTO. Image-free navigation tool [12] and kinematic navigation system such as Orthopilot [13,14] also exist. However, such approaches increase the time of the procedure [15], which may increase the risk of infection [15] and some are radiative for the patient and the surgeon. Moreover, errors due to mis-registration [15] or complications due to pin insertion [16] can occur. Finally, none of them can be used to provide information on the mechanical properties of the bone surrounding the osteotome.

Over the last few decades, vibroacoustic methods have been widely developed to estimate the stability of teeth [17], dental implants [18] and orthopedic implants, focusing on vibration response analysis [19] or impact sound analysis [20–22]. Our group has developed an impact method based on the use of a hammer instrumented with a force sensor located on the impacting face to assess implant stability during a Total Hip Arthroplasty (THA) surgery. This method has been used to assess the acetabular cup implant stability in an animal model [23] and in anatomical model [24] and has been validated through the development of numerical models [25]. The same conclusions were obtained for the femoral stem stability *in silico* [26], in artificial bones [27], in an animal model [28] and in anatomical model [29]. The force sensor enables to record a signal corresponding to the variation of the force as a function of time during each impact between the hammer and the osteotome. The method uses a dedicated signal processing technique and has been shown to provide information on the mechanical properties of the impacted system. Based on the results obtained in the context of THA, the method has been adapted to osteotomies since surgeons also use a surgical mallet. A preliminary study [30] on various materials has demonstrated that the method can be used to estimate the thickness and the mechanical properties of a sample being osteotomized. Other studies were carried out in the context of rhinoplasty *ex vivo* [31] and with anatomical subjects [32] to demonstrate that the instrumented hammer can detect crack initiation and transition from one bone to another. More recently, we have shown that the instrumented hammer can be used to predict the rupture of trabecular bone and plywood plate-like samples [33] as well as the rupture of the pterygoid plate *ex vivo* [34]. However, the method had never been employed to investigate the transition from cortical to trabecular bone (and *vice versa*) during an osteotomy.

The aim of the present *ex vivo* study is to determine whether an instrumented hammer can be used to assess when the material surrounding the tip of the osteotome changes (from cortical to trabecular bone and *vice versa*). To do so, fifteen osteotomies were carried out until rupture in fifteen pig tibia using the instrumented hammer. A dedicated signal processing technique was applied to each signal of each osteotomy. We developed an algorithm to detect the transitions from the cortical bone to the trabecular bone and from the trabecular bone to the cortical bone. The progression of the osteotome was monitored using a video camera. The results of the algorithm were compared with the progression of the osteotome and with the surgeon proprioception.

**2. Materials and methods**

**2.1 Experimental Procedure**

Fifteen pig legs were obtained from the local butcher shop. Each tibia was separated from the femur and the fibula. All the soft tissues around the tibia were removed to allow the visualization of the progression of the osteotome. The experiments were made in accordance with the guidelines of the ethical committee of the University Paris-Est Creteil (UPEC).

Each bone sample was placed into two 3D-printed seats (see Fig. 2) in order to stabilize their position during the impacts. These two seats were adapted to the shape of the proximal and distal parts of the tibia and could be moved along the *x*-axis to adapt to the bone length.

The osteotome used had a 3-mm straight edge (CP304-3, Microfrance, France). It was initially placed in the proximal part of the tibia and in the lateral region, in a plane perpendicular to the bone axis. When impacted, such positioning allows the osteotome to cross the lateral cortical bone, trabecular bone and eventually the medial cortical bone. The experiments were carried out by an experienced surgeon who provided an empirical feedback on the impact number corresponding to the transition between i) the lateral cortical bone and the trabecular bone, noted $P_t$ and ii) the trabecular bone and the medial cortical bone, noted $P_c$. To be as close as possible of the clinical situation, the hammer and the osteotome were held manually by the experienced orthopedic surgeon. The osteotome was impacted until its tip went out of the medial cortical bone and the total number of impacts realized during the osteotomy procedure for each bone sample (and thus each osteotomy) *#j* was noted *Q(j)*.

The mallet used was a classical surgical mallet (32-6906-26, Zepf, Tuttlingen, Germany) weighing 260 g on which a piezo-electric force sensor (208C05, PCB Piezotronics, Depew, NY, USA) was screwed (see Fig. 2). This sensor had a measurement range up to 35.59 kN in compression. A data acquisition module (NI 9234, National Instruments, Austin, TX, USA) with a sampling frequency of 51.2 kHz and a resolution of 24 bits was used to record the variation of the force as a function of time for a duration of 20 ms. Two signals *s(t)* corresponding to impacts of the hammer on the osteotome having its tip located in trabecular and cortical bone are shown in Fig. 3. Similarly, as in previous studies [30], the time of the maximum of the first two peaks of *s(t)* was determined for each impact. The indicator $\tau$ was then defined as the difference between the times of the second and first peaks of *s(t)*.

### 2.2 Determination of the material surrounding the osteotome tip

#### *2.2.1 Measure of the sinking of the osteotome*

For each impact #$i$ ∈ [1; $Q(j)$] corresponding to each osteotomy #$j$ ∈ [1; 15], the osteotome sinking *L(i,j)* was measured during the insertion, with *L=0* corresponding to its initial position before any impact (*i=0*). To do so, a millimeter graduation has been engraved using a laser on the osteotome side (see Fig. 4). A video camera (L-920M3, Spedal, Taiwan) was used to monitor the sinking of the osteotome into the bone after each impact.

#### *2.2.2 Determination of the type of bone around the osteotome tip*

Once the osteotomy was performed, each bone sample was cut in a plane corresponding to that of the osteotome. Each sample was then colored using Picrofuschin solution (1001990500, Sigma-Aldrich, Saint-Louis, Missouri, USA) in order to better distinguish trabecular from cortical bone. A microscope (Stemi 305, Carl Zeiss, Germany) was used to take a picture of regions of interest corresponding to the intersection of the two cortical bone layers with the osteotome path, as shown in Fig. 4. The intersection between cortical and trabecular bone was then segmented using Tracker (Tracker 6.1.6, National Science Foundation, Alexandria, VA, USA). The thickness of the first (respectively second) cortical bone layer was noted $D_{c1}$ (respectively $D_{c2}$). Both values correspond to the average of the three measurements taken along the osteotome width for each cortical layer, as illustrated in Fig. 4. For each cortical layers of all samples, one measurement was taken on each side of the osteotomy path (where the cortical layers were considered as intact) and another measurement was taken in the middle of the osteotome width. The trabecular bone thickness, $D_t$, was then evaluated thanks to an estimation of the external bone diameter $D_{total}$ obtained with a digital caliper following:

$$D_t = D_{total} - D_{c1} - D_{c2} \tag{1}$$

The type of bone surrounding the osteotome tip was determined by combining the information related to the sinking of the osteotome and to the values of the parameters $D_t$, $D_{c1}$ and $D_{c2}$. Each impact #$i$ ∈ [1; $Q(j)$] corresponding to each osteotomy #$j$ ∈ [1; 15] was classified into three categories based on the type of bone in which the osteotome tip was located:

- First cortical layer when:

$$L\ (i, j) \leq D_{c1}\ (j) \tag{2}$$

- Trabecular bone when:

$$D_{c1}(j) < L\ (i, j) \leq D_{c1}(j) + D_t(j) \tag{3}$$

- Second cortical layer when:

$$D_{c1}(j) + D_t(j) < L\ (i, j) \tag{4}$$

For each osteotomy #$j$, the first impact number when the osteotome tip reaches trabecular bone (respectively second cortical bone) was noted $N_t(j)$ (respectively $N_c(j)$).

### 2.2.3 Algorithm to detect transitions between cortical and trabecular bones

A dedicated algorithm was developed in order to detect the transitions between i) the first cortical bone layer and trabecular bone and ii) trabecular bone and the second cortical layer, based on the results obtained with the instrumented hammer and namely on the variations of $\tau$ as a function of $i$. Based on the results obtained in [30], an increase (respectively decrease) of $\tau$ is expected when the osteotome tip goes from cortical bone to the trabecular bone (respectively from the trabecular bone to the cortical bone).

For each osteotomy $\#j \in [1; 15]$, the osteotome tip was considered to reach the trabecular bone, according to the algorithm, for the first impact $\#M_t(j)$ respecting:

$$M_t(j) > 2 \quad \text{and} \quad \tau(M_t(j)) > k_t * (\tau(M_t(j) - 1) + \tau(M_t(j) - 2))/2 \qquad (5)$$

where $k_t$ is a constant to be determined using an optimization algorithm (see below).

Then, once the osteotome tip was considered to be located in trabecular bone, it was considered to reach the second cortical bone layer for the first impact $\#M_c(j)$ respecting:

$$\tau(M_c(j)) < k_c * (\tau(M_c(j) - 1) + \tau(M_c(j) - 2))/2 \qquad (6)$$

where $k_c$ is a constant to be determined using another optimization algorithm (see below).

The optimal values of $k_c$ and $k_t$ were determined following the procedure described below, which corresponds to a minimization under constraints algorithm. The goal was to determine the optimal value of $k_c$ (respectively $k_t$) allowing a minimization of the discrepancy $\tilde{\Delta}_c(k_c)$ (respectively $\tilde{\Delta}_t(k_t)$), which corresponds to the average value of $|M_c(j) - N_c(j)|$ (respectively $|M_t(j) - N_t(j)|$) for #$j \in [1; 15]$. An additional constraint was imposed, which corresponds to avoiding obtaining any sample for which $M_c(j) > N_c(j)$ (respectively $M_t(j) > N_t(j)$), in order to make sure that the algorithm always detects the transition before it occurs. The constraint of avoiding having $M_c(j) > N_c(j)$ (respectively $M_t(j) > N_t(j)$) for all samples was checked by imposing $B_t(k_t)=0$ and $B_c(k_c)=0$, where $B_t(k_t)$ (respectively $B_c(k_c)$) is the number of samples for which $M_t(j) > N_t(j)$ (respectively $M_c(j) > N_c(j)$) when considering the parameter $k_t$ (respectively $k_c$).

The algorithm aiming at determining the optimal values of $k_t$ and $k_c$ was thus composed of two parts.

The first part consists of determining the values of $M_t$ and then $M_c$ corresponding to two given values of $k_t$ and $k_c$. For each sample, the aim is to determine i) the first impact $M_t$ for which Eq. 5 is respected and ii) the first impact $M_c$ for which Eq. 6 is respected. Then, the values of $|M_t - N_t|$ and of $|M_c - N_c|$ corresponding to the difference between the impact number predicted by the instrumented hammer and by the video for each sample were determined.

The second part of the algorithm corresponds to the minimization procedure allowing to determine the optimal values of $k_t$ and $k_c$ minimizing the average value $\tilde{\Delta}_t(k_t)$ and $\tilde{\Delta}_c(k_c)$ of $|M_t - N_t|$ and of $|M_c - N_c|$ for all samples under the constraint that $N_t > M_t$ and $N_c > M_c$, respectively. This minimization procedure was simply done by analyzing the results shown in Fig. 6.

## 3. Results

### 3.1 Statistics on the impacts features

Table 1 shows the statistical parameters obtained for the thickness of both cortical layers and of the trabecular length.

For the fifteen osteotomies, a total number of 540 impacts were obtained with, according to the procedure described in subsection 2.2, 134 impacts in the first cortical layer, 188 in the second cortical layer and 218 in trabecular bone. We obtained an average number of 34.9 ± 11.7 impacts per osteotomy, distributed as indicated in Table 2.

### 3.2 Behavior of the indicator $\tau$

Figure 5 shows the variation of $\tau$ as a function of the impact number during the osteotomy procedure for the sample #9. Three regimes may be distinguished. First, the value of $\tau$ is approximately constant around 0.530 ± 0.059 ms, which corresponds to a regime where the osteotome tip is located in the first cortical layer. Then, $\tau$ quickly increases to reach a mean value of 1.118 ± 0.236 ms, which corresponds to a regime where the osteotome tip is located in trabecular bone. Eventually, $\tau$ decreases to a mean value of 0.484 ± 0.041 ms, which corresponds to a regime where the osteotome tip is located in the second cortical layer.

Statistical parameters corresponding to the values of $\tau$ obtained for the 15 osteotomies are shown in Table 3. The mean values of $\tau$ obtained in trabecular bone are significantly higher than in both cortical layers, which has been shown following an ANOVA analysis ($p < 0.001$, $F$=215.02).

### 3.3 Algorithm for the instrumented hammer

Figure 6A (respectively 6B) shows the variation of $\tilde{\Delta}_t$ (respectively $\tilde{\Delta}_c$) and of $B_t$ (respectively $B_c$) as a function of $k_t$ (respectively $k_c$). In both cases, $\tilde{\Delta}_t$ and $\tilde{\Delta}_c$ are shown to exhibit a local minimum value, while $B_t$ and $B_c$ both increase as a function of $k_t$ and $k_c$, respectively. The algorithm described in subsection 2.2.3 leads to: $k_t$ = 1.159 and $k_c$= 0.699 and these values will be used in what follows.

### 3.4 Comparison between the different methods

Figure 7A shows the distribution, for all 15 osteotomy procedures, of the values of:

- $N_t$ - $P_t$, in light grey, which corresponds to the difference of the impact number predicted by the video and by the surgeon for the transition between the first cortical bone layer and trabecular bone

- $N_t$ - $M_t$, in black, which corresponds to the difference of the impact number predicted by the video and by the instrumented hammer for the transition between the first cortical bone layer and trabecular bone

- $M_t$ - $P_t$, in dark grey, which corresponds to the difference of the impact number predicted by the instrumented hammer and by the surgeon for the transition between the first cortical bone layer and trabecular bone.

A perfect agreement is obtained between the impact numbers predicted by the video and the surgeon (respectively the instrumented hammer) for eight samples, *i.e.* 53 % (respectively in 9 samples, *i.e.* 60 %) and the difference is within ± 1 impacts for 14 samples, *i.e.* 93.3 % (respectively 11 samples, *i.e.* 73.3 %).

The value of $N_t$ - $M_t$ was always positive (a condition imposed by $B_t(k_t)$=0), which indicates that the instrumented hammer always predicts the arrival of the osteotome in trabecular bone before the video.

Except in one case for which $M_t(4)$- $P_t(4)$= 1, the value of $M_t(j)$- $P_t(j)$ was always negative or equal to zero, which indicates that the instrumented hammer could almost always predict the arrival of the osteotome in trabecular bone at worst at the same impact as the surgeon. These results emphasize the interest of the present method to assist the surgeon during the osteotomy procedure.

Similarly, Fig. 7B shows the distribution, for all 15 osteotomy procedures, of the values of:

- $N_c$ - $P_c$, in light grey, which corresponds to the difference of the impact number predicted by the video and by the surgeon for the transition between the trabecular bone and the second cortical bone layer

- $N_c$ - $M_c$, in black, which corresponds to the difference of the impact number predicted by the video and by the instrumented hammer for the transition between the trabecular bone and the second cortical bone layer

- $M_c$ - $P_c$, in dark grey, which corresponds to the difference of the impact number predicted by the instrumented hammer and by the surgeon for the transition between the trabecular bone and the second cortical bone layer.

The value of $N_c$ - $M_c$ was always positive, this condition was imposed by $B_c(k_c)$=0 but it also corresponds to the minimum value of $\tilde{\Delta}_c$. This result indicates that the instrumented hammer could always predict the arrival of the osteotome in cortical bone before the video. The early prediction of the instrumented hammer may be explained by the higher mechanical properties of the homogenized medium surrounding the osteotome when it approaches cortical bone.

A similar behavior is obtained for $N_c$ - $P_c$ (all values are also positive), which indicates that the surgeon could detect the arrival on cortical bone before the video.

Eventually, the difference between the detection of the arrival on cortical bone predicted by the instrumented hammer and by the surgeon is within ± 1 impacts for 13 samples (87% of the cases). In one case, the instrumented hammer predicts the arrival on cortical bone two impacts after the surgeon, while for another case, the instrumented hammer predicts the arrival on cortical bone three impacts before the surgeon.

The statistical values (average and standard deviation) of $|N - P|$, $|N - M|$ and $|M - P|$ for the transitions between the first cortical bone layer and the trabecular bone and between the trabecular bone and the second cortical bone layer are shown in Table 4.

## 4. Discussion

The main originality of this study was to investigate whether an instrumented hammer could be used to detect changes of materials surrounding the osteotome tip during an osteotomy. The results show the potentialities of such device to provide a decision-support system to the surgeon during osteotomies such as the HTO. A criteria based on the variation of $\tau$ was validated in order to determine the impact number corresponding to the transition between the first cortical bone layer and trabecular bone and between the trabecular bone and the second cortical bone layer.

### 4.1 Physical explanation

A physical interpretation of the behaviour of $\tau$ had already been given in [30], where a simple analytical model was able to predict signals corresponding to the variation of the force as a function of time during an impact that were qualitatively similar to the ones shown in Fig. 3. We showed that the different peaks of the force result from the rebound of the osteotome between bone tissue and the head of the hammer during the impact. Therefore, the time between the first two peaks (given by $\tau$), is related to the rigidity of the material around the osteotome tip. As shown in Table 3 and Fig. 5, $\tau$ increases significantly when the tip of the osteotome is located in trabecular bone compared to cortical bone, because of the higher rigidity of cortical bone compared to trabecular bone.

### 4.2 Comparison of the performances of the different methods

As indicated in Fig. 7 and Table 4, there are several differences in the prediction of the two transitions between the three methods, which may be explained by the various sources of errors associated to each method.

First, while the orthopedic surgeons were experienced, their proprioception is not perfectly accurate and potential errors are always possible because it is an empirical approach. However, it remains difficult to precisely quantify the error corresponding to the estimation of $P_t$ and $P_c$.

Second, different errors are associated to the video analysis. In particular, the measurement of the osteotome sinking has an error of the order of 0.5 mm, due to the precision of the graduation. Moreover, due to the irregular shape of the cortical layer, the three measurements performed to assess the value of each cortical layer (see section 2.2.2 and Fig. 4) had an averaged standard deviation of 0.06 mm and 0.08 mm for $D_{c1}$ and $D_{c2}$, respectively. The error corresponding to the estimation of $D_{total}$ was of the order of 0.03 mm, which corresponds to the incertitude of the digital caliper. As a consequence, the error corresponding to the estimation of the osteotome is driven by the incertitude on the measurement of the sinking, which is equal to around 0.5 mm. A sinking of 0.5 mm typically corresponds to a number of impacts equal to around 0.66 ± 0.24 impacts, which provide an error on the estimation of $N_t$ and $N_c$ with the video.

A relatively good agreement is obtained between the estimation of the video and of the surgeon, which may be explained by the fact that the experimental conditions are optimal compared to the clinical situation. In particular, there is no surrounding noise, the surgeon has time to perform the osteotomy without any stress, there is no soft tissue to affect the proprioception and the operated area is unobstructed so that the surgeon has visual access to the surgical site. Moreover, the surgeon performing the osteotomy is a very experienced surgeon.

Third, there are several errors related to the instrumented hammer, which are associated with the method presented in this work and to the complex nature of bone tissue. It is precisely one of the aims of the present paper to assess the error corresponding to the estimation of $M_t$ and $M_c$.

Despite the aforementioned discrepancies between the three methods, the instrumented hammer was always able to predict the transition between trabecular bone and cortical bone early enough to avoid a complete rupture of the sample. This performance will be a key point to prevent this severe complication of HTO if such a decision support-system is to be used in a clinical environment in the future.

### 4.3 Limitations

This study has several limitations. First, there are significant differences between the clinical situation and the present animal model, in particular in terms of general configuration (fixation of the sample in a vice), of bone properties and of the presence of soft tissues. Note that the presence of soft tissues was shown not to affect the results in a previous paper [35] considering the use of the instrumented hammer in total hip replacement, which needs however to be confirmed in the context of HTO. We chose to work *ex vivo* considering pig tibia because it is a widely used model for HTO [36,37] and to test new HTO fixation devices [38,39]. The estimated Young's modulus of pig cortical bone (19.4 ± 5.8 GPa according to [40]) is quite similar to the human one (ranging from 18 up to 28 GPa according to [41–43]). Regarding the trabecular bone, the Young's modulus of the pig trabecular bone (0.073 ± 0.015 GPa according to [44]) is in the low range of the human one (ranging from 0.01 up to 14.13 GPa according to [41,45,46]). The choice of considering *ex vivo* animal model is also justified by the necessity of validating the method before considering a study with anatomical subjects, which is related to ethical issues.

Second, the instrumented hammer also has limitation related to the fact that this technique is new and currently under development. In particular, only 15 samples were considered and further work should focus on a more important number of samples. Due to the relatively low number of sample considered, it remains difficult to consider the impact of using a limited dataset for parameter determination, followed by validation with the remaining data, which would be interesting. While more experimental data are needed to conclude on the reliability of the method, this study is the first one regarding the use of the hammer for HTO. Since an animal study is required for the application of the technique to anatomical subjects, we choose to perform a small feasibility study with a few samples and then to move to anatomical subjects. Note that a similar approach considering 14 (respectively 22) biological samples has already been applied in [31] (respectively [33]). Moreover, we chose a relatively simple criteria based on the last two impacts and other options such as taking into account more impacts could have been considered. Choosing a criteria based on more impacts was impossible because the minimum number of impacts in the first cortical bone layers is three (see Table 2). Therefore, choosing a criteria taking in account more impacts would have automatically resulted in a delay ($B_t > 0$).

Third, only one type of osteotome was used, while several other osteotomes are on the market, which may affect the result. The size and shape of the osteotome may modify the signals obtained with the instrumented hammer, which explains why other osteotome geometry should also be tested in the future. The values of $\tau$ have been related to the resonance frequency of the system formed by the osteotome and bone tissue in contact with its tip [30]. Therefore, the values of $\tau$ are likely to be modified if another osteotome is used. However, the principle of the measurement is similar for any osteotome, since the values of $\tau$ are also related to the nature of the bone in contact with its tip. Therefore, while a similar variation is expected to $\tau$ when the osteotome tip arrives in cortical or trabecular bone, the values of $k_t$ and $k_c$ may depend on the osteotome and should be determined in a large dataset obtained in anatomical subjects and then in a clinical study. Note that we assumed that osteotome blade was in contact with the same material, which is necessary to exploit the data.

 Fourth, adding another sensor to the system could provide interesting information such as the frequency response function, similarly as what was done in the context of THA [20,47] where sensors were placed on the ancillary or on the implant. However, we chose not to apply this approach herein because it complicates the experimental set-up in the context of a future clinical use. Moreover, we have already shown in [30] that the instrumented hammer allows to measure stiffness variations of the impacted material independently of the impacting force. Note that here, we did not recommend any impacting force, which corresponds to the future clinical situation.

## 5. Conclusion

This *ex vivo* animal study investigates the performances of an instrumented hammer to determine the position of the osteotome tip and namely its transition from trabecular to cortical bone and *vice versa*. This situation is of particular interest in the case of HTO, because of the importance of keeping the hinge intact and thus of stopping impacting the osteotome at the right moment. This detection is particularly interesting for HTO to detect the transition from trabecular to cortical bone because it corresponds to the moment where impaction must end. Our approach could therefore prevent unnecessary impacts leading to potential complications. Future studies should be performed with anatomical subject to be as close as possible to the anatomical and clinical situation.

**FUNDING**

This project has received funding from the project OrthAncil (ANR-21- CE19-0035-03), from the project OrthoMat (ANR-21-CE17-0004) and from the project ModyBe (ANR-23-CE45-0011-02). This project has received funding from the European Research Council (ERC) under Horizon 2020 (grant agreement # 101062467, project ERC Proof of Concept Impactor).

**Figure List**

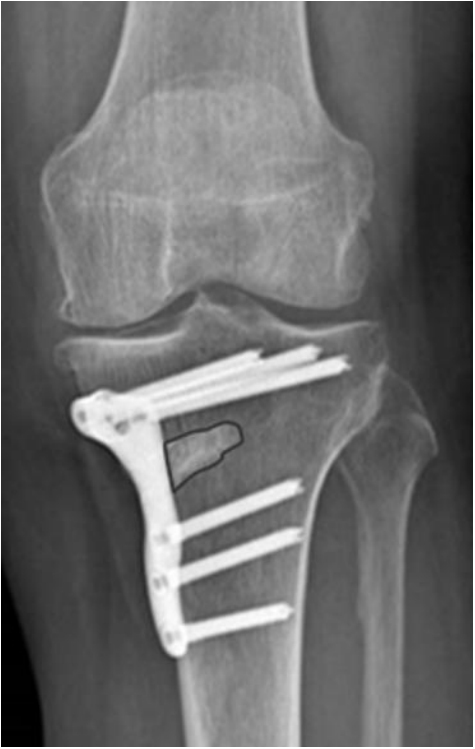

Fig. 1          Postoperative X-ray imaging from a patient who underwent a HTO. Plate

and screws used to maintain the bone integrity are shown as well as the bone filler surrounded by a red line.

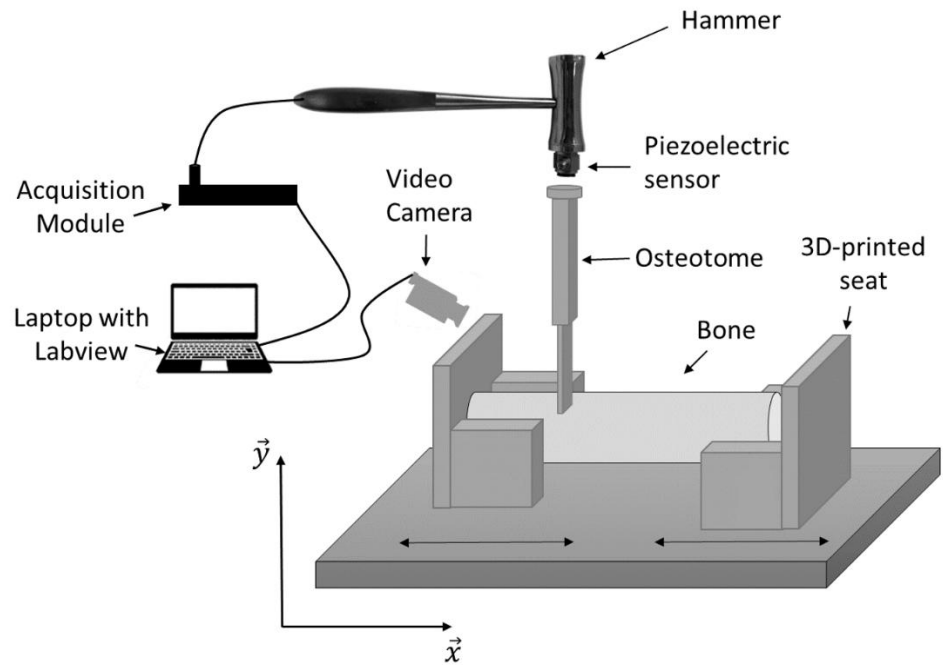

Fig. 2    Schematic representation of the experimental configuration showing the pig bone sample placed in the 3D-printed seat, the osteotome and the piezoelectric sensor. The location of the video camera is also shown.

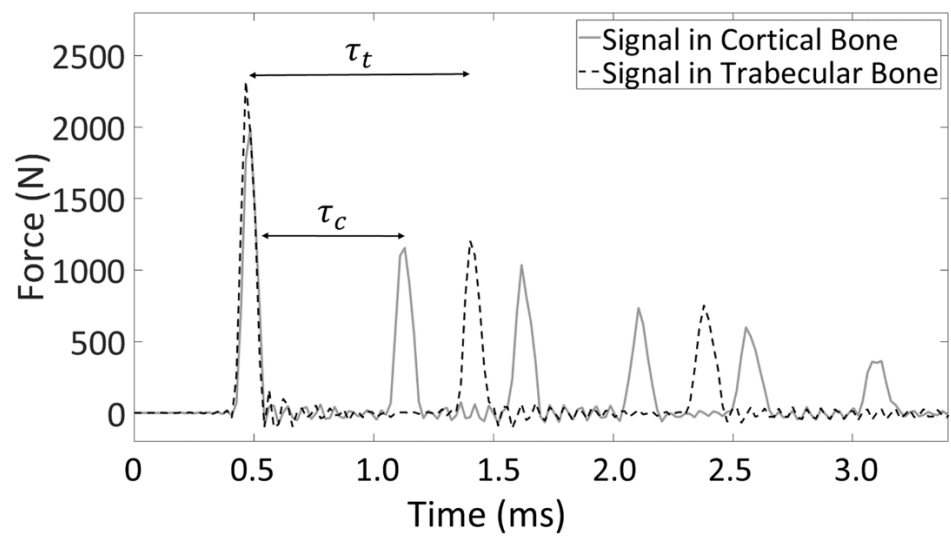

Fig. 3    Example of two signals *s(t)* corresponding to the variation of the force as a function of time obtained for an impact of the instrumented hammer on the osteotome having its tip located in trabecular (dashed black line) and cortical bone (solid gray line). $\tau_t$ (respectively $\tau_c$) denotes the time between the two first peaks of the signal for the osteotome having its tip located in trabecular (respectively cortical) bone.

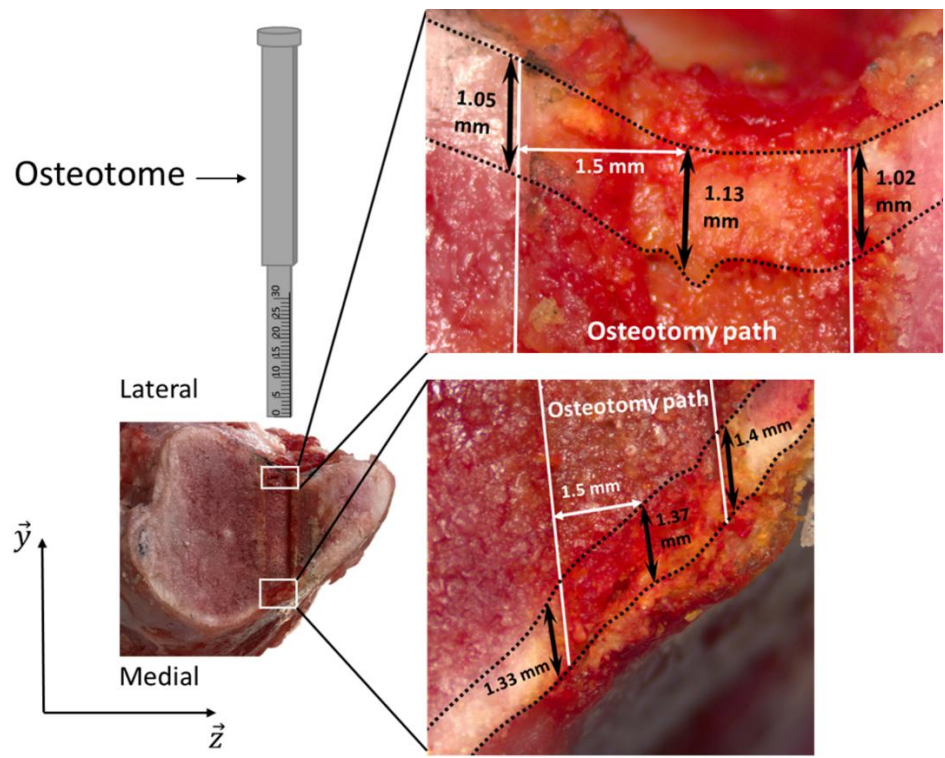

Fig. 4    Representation of the engraved osteotome and of the sample after it was cut to assess the thickness of the first and second cortical bone layers. The cortical layers are indicated by the black dotted lines. The osteotomy path is indicated by the white full lines.

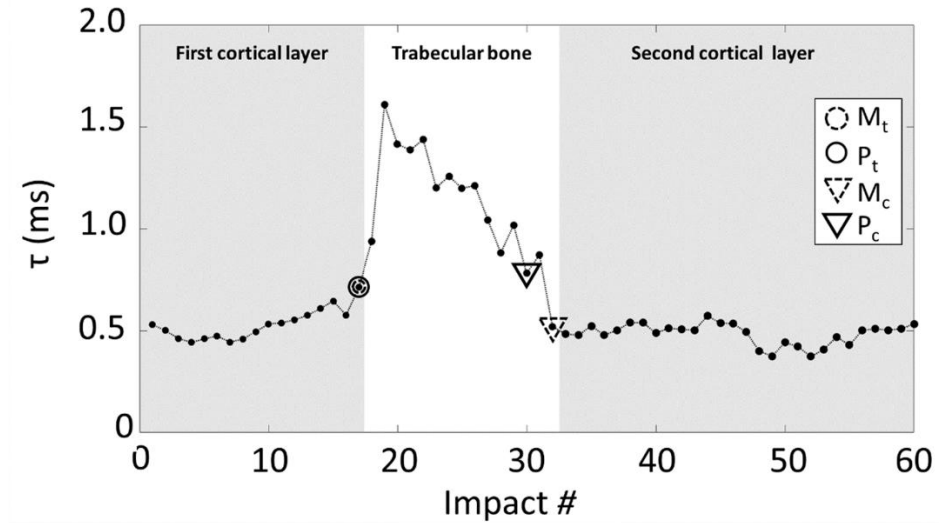

Fig. 5    Variation of $\tau$ as a function of the impact number for the sample #9. The gray areas correspond to impact numbers where the osteotome tip is located in both cortical layers according to video analysis. Circles (respectively triangles) correspond to the transition from the first cortical bone layer to the trabecular bone (respectively to the transition from the trabecular bone to the second cortical bone layer). The solid (respectively dashed) triangles and circles correspond to the detection $P$ by the surgeon (respectively $M$ by the instrumented hammer with the algorithm after optimization, i.e. for $k_t$= 1.159 and $k_c$=0.699).

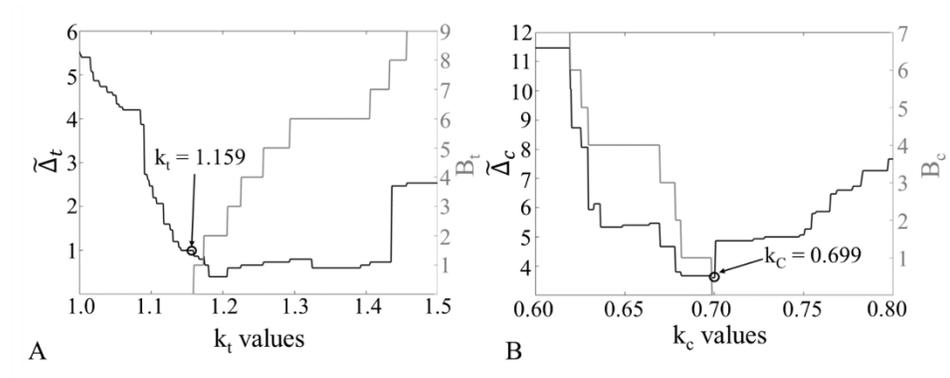

Fig. 6    Results used to find the optimal value of $k_t$ and $k_c$. A: Variation of the discrepancy $\tilde{\Delta}_t$ and of the number of samples $B_t$ for which $M_t(j) > N_t(j)$ as a function of $k_t$. B: Variation of the discrepancy $\tilde{\Delta}_c$ and of the number of samples $B_c$ for which $M_c(j) > N_c(j)$. as a function of $k_c$.

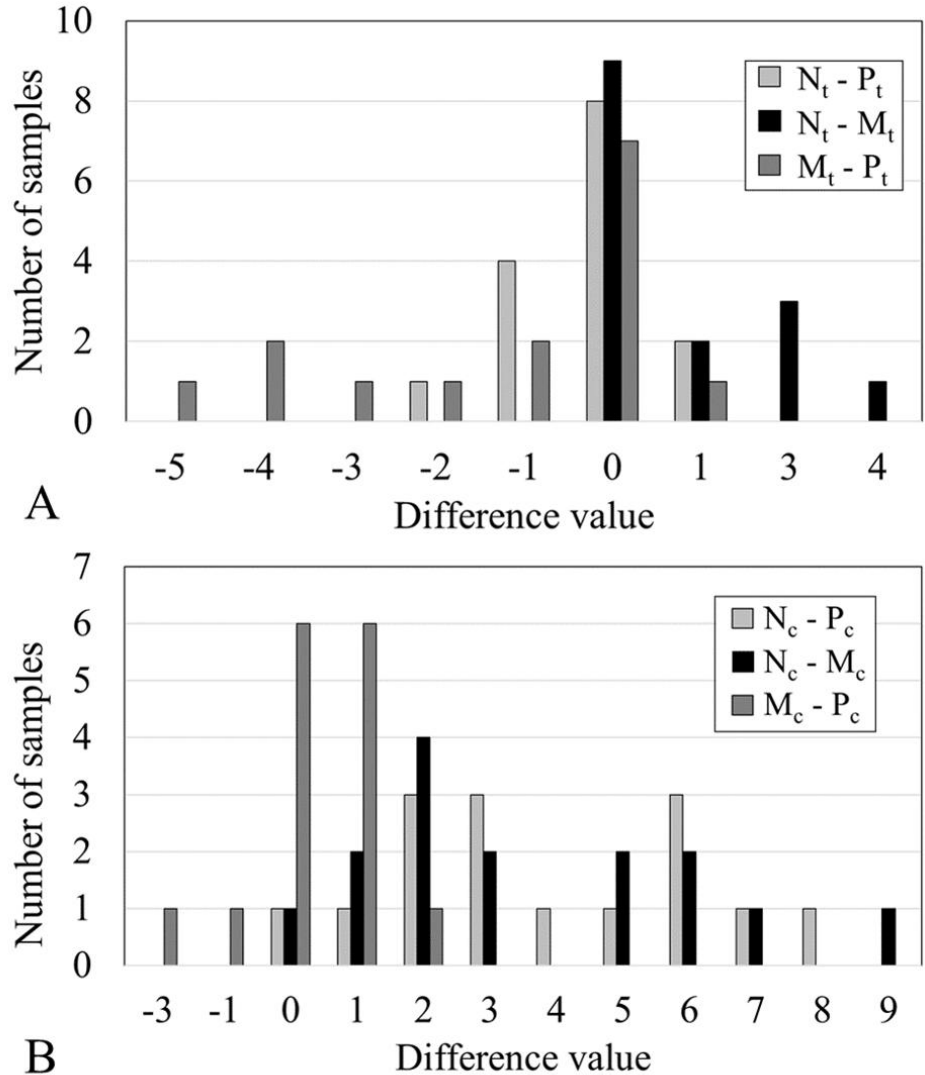

Fig. 7      A: transition between the first cortical bone layer and trabecular bone.

Distribution of the difference of the impact number predicted by the video

($N_t$) and by the surgeon ($P_t$), of the impact number predicted by the video

($N_t$) and by the instrumented hammer ($M_t$) and of the impact number

predicted by the instrumented hammer ($M_t$) and the surgeon ($P_t$). B:

transition between trabecular bone and the first cortical bone layer.

Distribution of the difference of the impact number predicted by the video

($N_c$) and by the surgeon ($P_c$), of the impact number predicted by the video ($N_c$) and by the instrumented hammer ($M_c$) and of the impact number predicted by the instrumented hammer ($M_c$) and by the surgeon ($P_c$).

**Table Caption List**

Table 1    Statistical parameters on the thickness of the sample for all experimental samples

Table 2    Statistical parameters on the number of impacts per osteotomy for all experimental samples

Table 3    Average and standard deviation values of $\tau$ for all samples obtained in the first and second cortical layer and in trabecular bone, as determined by the sinking of the osteotome.

Table 4    Mean and standard deviation of the difference between the detection of the instrumented hammer ($M$), the surgeon ($P$) and the video camera ($N$).

Table 1

| Thickness of the: (mm) | Mean ± Standard deviation | Min | Max |
|---|---|---|---|
| First cortical layer | 1.24 ± 0.21 | 0.95 | 1.62 |
| Second cortical layer | 1.47± 0.37 | 1.05 | 2.28 |
| Total length | 18.00 ± 3.64 | 11.10 | 25.93 |
| Trabecular length | 15.28 ± 3.38 | 8.82 | 22.34 |

Table 2

| Number of impacts given in the: | Mean ± Standard deviation | Min | Max |
|---|---|---|---|
| First Cortical layer | 8.9 ± 4.1 | 3 | 26 |
| Trabecular Bone | 13.7 ± 4.3 | 5 | 31 |
| Second Cortical layer | 12.3 ± 4.8 | 2 | 28 |

Table 3

| $\tau$ (ms) | Mean ± Standard deviation | Min | Max |
|---|---|---|---|
| First Cortical layer | 0.579 ± 0.077 | 0.441 | 0.892 |
| Second Cortical layer | 0.484 ± 0.072 | 0.288 | 0.925 |
| Trabecular Bone | 0.969 ± 0.283 | 0.234 | 2.416 |

Table 4

| Transitions between | $\|N - P\|$ | $\|N - M\|$ | $\|M - P\|$ |
|---|---|---|---|
| 1st cortical bone layer and trabecular bone | $0.53 \pm 0.63$ | $1.0 \pm 1.5$ | $1.4 \pm 1.8$ |
| trabecular bone and the 2nd cortical bone layer | $3.8 \pm 2.4$ | $3.6 \pm 2.6$ | $0.8 \pm 0.9$ |